\documentclass{aastex}          
\usepackage{spr-astr-addons}    
\usepackage{graphicx}
\usepackage{wrapfig}
\usepackage{amsmath}
\usepackage{amssymb}
\usepackage{bm}
\usepackage{latexsym}
\usepackage{color}

\begin{document}

\title{The case for inflow of the broad-line region of active galactic nuclei}
\slugcomment{To appear in {\it Spectral Line Shapes in Astrophysics},  Astrophysics \& Space Science.}
\shorttitle{The case for inflow of the broad-line region}
\shortauthors{Gaskell \& Goosmann}

\author{C. Martin Gaskell}
\affil{Department of Astronomy and Astrophysics, University of California at Santa Cruz, Santa Cruz, CA 95064, USA}
\and \author{Ren\'e W. Goosmann}
\affil{Observatoire astronomique de Strasbourg, Université de Strasbourg, CNRS, UMR 7550, 11 rue de l'Universit\'e, F-67000 Strasbourg, France}

\email{mgaskell@ucsc.edu}

\begin{abstract}
The high-ionization lines of the broad-line region (BLR) of thermal active galactic nuclei (AGNs) show blueshifts of a few hundred km/s to several thousand km/sec with respect to the low-ionization lines.  This has long been thought to be due to the high-ionization lines of the BLR arising in a wind of which the far side of the outflow is blocked from our view by the accretion disc.  Evidence for and against the disc-wind model is discussed.  The biggest problem for the model is that velocity-resolved reverberation mapping repeatedly fails to show the expected kinematic signature of outflow of the BLR.  The disc-wind model also cannot readily reproduce the red side of the line profiles of high-ionization lines.  The rapidly falling density in an outflow makes it difficult to obtain high equivalent widths.  We point out a number of major problems with associating the BLR with the outflows producing broad absorption lines.  An explanation which avoids all these problems and satisfies the constraints of both the line profiles and velocity-resolved reverberation-mapping is a model in which the blueshifting is due to scattering off material spiraling inwards with an inflow velocity of half the velocity of the blueshifting.  We discuss how recent reverberation mapping results are consistent with the scattering-plus-inflow model but do not support a disc-wind model.  We propose that the anti-correlation of the apparent redshifting of H$\beta$ with the blueshifting of \ion{C}{4} is a consequence of contamination of the red wings of H$\beta$ by the broad wings of [\ion{O}{3}].
\end{abstract}



\section{Introduction}
\label{sec:intro}

One of the most basic questions to be answered in trying to understand the processes dominating motions of gas close to a supermassive black hole is, ``how is the gas moving?''  In thermal active galactic nuclei\footnote{Thermal AGNs are high accretion rate AGNs where the energy output is dominated by thermal emission of the accretion disc.  Non-thermal AGNs, which we do not consider here, are very low accretion rate AGNs where the energy output is dominated by the mechanical energy of the radio jet.  See \citet{Antonucci12} for detailed discussion.} (AGNs) much of this gas can be seen as the broad-line region (BLR) which produces the characteristic broad emission lines of thermal AGNs.  Two important observational constraints that a model of the kinematics of the BLR has to explain are the {\em profiles} of various emission lines and the {\em velocity-dependent time delays} in the responses of the lines to continuum variability.  A successful model must also explain the equivalent widths of the lines and the line intensity ratios as a function of velocity.

From the profiles a key phenomenon to be explained is the systematic blueshifting of the high-ionization broad lines with respect to the systemic velocity of the host galaxy \citep{Gaskell82}. In this contribution we review and discuss the two main models proposed to explain the blueshifting. As we shall see, these models require fundamentally different kinematics so it is clearly important to know the cause of the blueshifting in order to understand how thermal AGNs work.

\section{Observational properties}

From a comparison of the wavelengths of low- and high-ionization broad lines in active galactic nuclei (AGNs) \citet{Gaskell82} discovered that, unlike low-ionization lines, the high-ionization broad lines of AGNs have a systematic asymmetry.  The centroids of the broad lines were blueshifted by about 600 km s$^{-1}$ on average with respect to the low-ionization lines.  Comparison of the wavelengths of low-ionization broad lines with the wavelengths of narrow lines showed that it was the high-ionization lines that are blueshifted rather than the low-ionization lines being redshifted. Typical profile shifts are shown in Figure 1. From these it can be seen that the blueshifting can lead to a huge asymmetry in the most extreme cases.

\begin{figure}[t]
 \vspace{5pt}
 \centering \includegraphics[width=8cm]{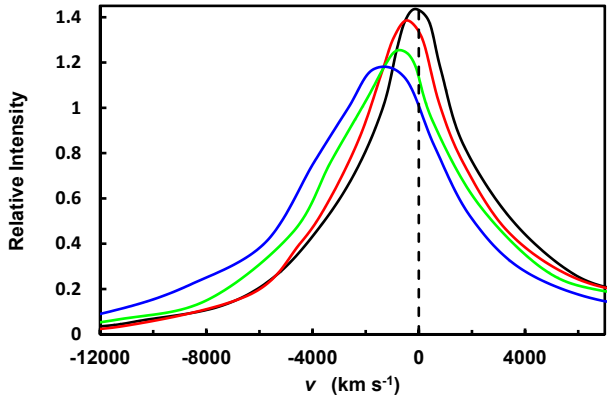}
\caption{Shifts in composite SDSS \ion{C}{4} profiles binned by blueshift.  The curves are normalized to the same area.  The blueshifts of each sample
are 197, 606, 1003, and 1526 km s$^{-1}$ for the black, red, green, and blue curves respectively. (Figure from \citealt{Gaskell+Goosmann13}.)}
 \vspace{5pt}
\end{figure}

There have been many follow-on studies of the blueshifting effect.  \ion{C}{4} is not the only high-ionization line showing a blueshift.  \citet{Corbin90} and \citet{Steidel+Sargent91} found that there was a lesser blueshift of \ion{C}{3}] $\lambda$1990 and that the blueshifts of \ion{C}{4} and \ion{C}{3}] were correlated.  \citet{Tytler+Fan92} found that blueshifts of lines were roughly in order of increasing ionization.  \citet{Corbin90} found that the blueshifting of \ion{C}{4} was greatest for AGNs with a lower \ion{C}{4} equivalent width.  The most extreme examples are the so-called ``weak-line quasars'' (WLQs) most of which show large blueshifts of \ion{C}{4} (e.g., \citealt{Wu+12}). \citet{Corbin90} also found that broad absorption line objects (BALQSOs) showed unusually large blueshifts. \citet{Steidel+Sargent91} and \citet{Corbin92} found that the blueshift of \ion{C}{4} was about three times greater for radio-quiet AGNs than for radio-loud AGNs.  These studies of \citet{Corbin90,Corbin92} also showed that the blueshifting was greater for higher luminosity AGNs. These correlations have been confirmed using larger samples of AGNs \citep{Tytler+Fan92,Richards+02}.

\citet{Sulentic+00} showed that the blueshifting of \ion{C}{4} is strongest in AGNs with high Eddington ratios, $L/L_{\textrm{Edd}}$, where $L_{\textrm{Edd}}$  is the Eddington luminosity, which is proportional to the mass of the black hole, $M_\bullet$.  The AGNs with the highest Eddington ratios are the so-called ``narrow-line Seyfert 1s'' (NLS1s).  These also show the highest blueshifts (see \citealt{Leighly+Moore04}). The connection between relative accretion rate and the blueshifting of high-ionization broad lines is an important clue to the cause of the blueshifting.  Also interesting is the finding of \citet{Zamanov+02} that the blueshifting of the forbidden [\ion{O}{3}] lines is correlated with the blueshifting of \ion{C}{4}. This suggests either a causal connection between the BLR and the narrow-line region (NLR) or at least common factors, such as $L/L_{\textrm{Edd}}$, influencing both.

Two things the blueshifting of \ion{C}{4} does {\em not} correlate with are luminosity and redshift \citep{Richards+02}. This is important because it means that studies of low-luminosity AGNs (especially reverberation mapping) can be used to address the question of the cause of the blueshifting.

\section{The disc-wind model}

\citet{Gaskell82} proposed a ``disc-wind'' model as a possible explanation of the blueshifting.  Low-ionization lines would arise from gas associated with the accretion disc while the high-ionization lines would arise in an outflowing wind.  Since the accretion disc is optically thick, it blocks our view of the far side of the outflow and thus suppresses the red wing of the lines.  This gives a net blueshift.  This model is illustrated in the cartoon in Figure 2.
%
\begin{figure}[t]
 \vspace{5pt}
 \centering \includegraphics[width=7cm]{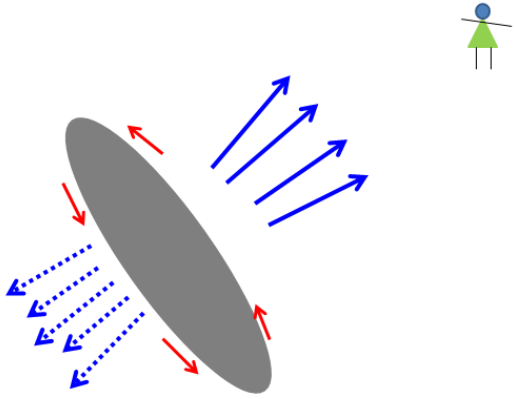}
 \caption{A cartoon illustrating the disc-wind explanation of the blueshifting of high-ionization lines.  In this model the low-ionization lines come from gas orbiting with the accretion disc (shown by red arrows) while the high-ionization lines arise an outflowing wind (shown as blue arrows).  The observer in the upper right has her view of the receding side of the outflow (shown as dotted lines) blocked by the accretion disc.}
\end{figure}

The disc-wind model obviously requires a BLR with two distinct components: one is a virialized, low-ionization component, and the other a high-ionization outflowing component (see \citealt{Gaskell87} and \citealt{Collin-Souffrin+Lasota88}).

\section{Strengths of the disc-wind model}

The outflow+obscuration (i.e., disc-wind) explanation of \citet{Gaskell82} has a number of strengths on theoretical and observational grounds and has therefore long been the favoured explanation.

\subsection{Outflows of material from AGNs are unambiguously detected through blueshifted absorption lines}

At the time of the discovery of the blueshifting of high-ionization broad emission lines the disc-wind model was well motivated because blueshifted broad {\em absorption} lines in BALQSOs were well known (see, for example, the review of \citealt{Weymann+81}). This absorption must necessarily arise in an outflow.  An additional motivation was that radiative acceleration of the broad-line region was thought probable because it could explain the ``logarithmic'' Balmer line profiles thought to be seen in most AGNs \citep{Blumenthal+Mathews75}.

\subsection{Theoretical expectations}

AGNs must have winds since a wind is necessary to remove angular momentum and allow accretion onto the black hole \citep{Blandford+Payne82}.\footnote{It is important to recognize that viscosity alone in an accretion disc, while important for energy generation, does not allow the accretion of material from infinity onto a black hole -- see Figure 1 of \citet{Pringle81}.}  The wind will be highly ionized and produce at least some emission in high-ionization lines.  It is thus natural to associate this with the blueshifted emission of high-ionization broad lines in AGNs.

\subsection{Analogy with the narrow-line region blueshifts}

Indirect support for the disc-wind explanation comes by analogy with the narrow-line region (NLR).  The [\ion{O}{3}] $\lambda\lambda$ 5007, 4959 lines show a modest systematic blueshift (a few hundred km s$^{-1}$) which \citealt{Heckman+81} proposed is a consequence of the combination of outflow and obscuration and suppression of the red side of the line profile by dust.  From spatially-resolved observations of the NLR there is no doubt that the inner, high-ionization NLR is outflowing (see \citealt{Crenshaw+10} and references therein).  \citet{Crenshaw+10} show that collapsing long-slit {\it Hubble Space Telescope} ({\it HST}) spectra of Seyfert galaxies to ground-based spatial resolution reproduces the observed, blueshifted [\ion{O}{3}] spectra.  The blueshifting of the high-ionization NLR lines is thus completely consistent with an outflow+obscuration model.

\subsection{The connection between BLR and NLR blueshifts}

Support for the NLR and high-ionization BLR blue-shifts having a similar mechanism comes from the correlation between them.  \citet{Zamanov+02} pointed out that the largest [\ion{O}{3}] blueshifts (shifts $< - 250$ km s$^{-1}$, which they refer to as ``blue outliers'') are found in sources where the broad high-ionization lines {\em also} show a large blueshift.  \citet{Zhang+11} show, for a much larger sample, that the blueshift of [\ion{O}{3}] is correlated with $L/L_{\textrm{Edd}}$ (see their Figure 4).  This is similar to the dependence of the broad-line blueshifting on $L/L_{\textrm{Edd}}$.

\section{Problems with the disc-wind model}

While the disc-wind model clearly has a lot going for it, it does have some major problems.  We detail here some of these problems.

\subsection{Inconsistency with velocity-resolved reverberation mapping results}

The foremost problem for the disc-wind model it that it is not consistent with velocity-resolved reverberation mapping.  The disc-wind model makes a definite prediction: if the high-ionization line emission is arising in a wind, then the blue side of the line profile preferentially arises from gas on the near side of the AGN. \citet{Gaskell88} pointed out that the direction of flow could be determined from velocity-resolved reverberation mapping.  The emission from the approaching near side in Figure 2 will respond almost instantaneously to continuum changes while the response of the receding outside will be delayed by twice the mean light-travel time to the gas.  Since the mean light-travel time is known from cross-correlating the variability of the whole line with the continuum \citep{Gaskell+Sparke86}, the outflow model leads to a definite prediction for each broad line in each AGN of how much the variability of the red side of a line should lag the continuum variability. As \citet{Gaskell88} showed (see Figure 3), the observed cross correlation functions of the wings of \ion{C}{4} of NGC~4151 do {\em not} match the prediction and instead imply gravitational domination of the gas motions.
\begin{figure}[t]
 \centering \includegraphics[width=8cm]{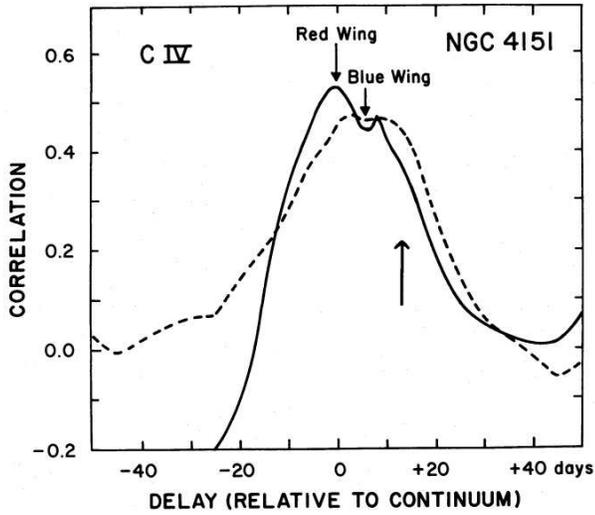}
 \caption{The cross correlation functions of the red wing (dashed line) and blue wing (solid line) of \ion{C}{4} $\lambda$1549 versus the continuum at $\lambda$1450 for NGC~4151.  A positive lag corresponds to the line emission following the continuum emission.  The lower arrow indicates the predicted peak of the cross correlation function of the red wing for pure outflow (i.e., if \ion{C}{4} originates in a wind). (Figure adapted from \citealt{Gaskell88}.)}
\end{figure}

It is important to recognize that {\em the lack of the predicted wind outflow kinematic signature is a general result}.  Identical results were subsequently obtained for Fairall 9 (compare Figure 3 here with Figure 10 of \citealt{Koratkar+Gaskell89}) and NGC 5548 (compare Figure 3 here with Figure 3 of \citealt{Crenshaw+Blackwell90}).  Re-observation of NGC~5548 gave the same result \citep{Korista+95,Done+Krolik96}.  Lower-quality UV observations of a number of other AGNs also failed to show evidence of outflow of \ion{C}{4} \citep{Koratkar+Gaskell91}.

\subsection{The BLR radius-velocity relationship}

Independent evidence for the gravitational domination of the motions of all BLR gas comes from the correlation of Doppler widths of lines, $\sigma$, with effective radii of emission as was first pointed out by \citet{Krolik+91}. The effective radius is given by the lag, $\tau$, in the response of a line to continuum variability \citep{Gaskell+Sparke86}.  If the gas motions are virialized then we should get $\sigma \propto \tau^{-1/2}$.  \citet{Krolik+91} obtained a best fit of $\sigma \propto \tau^{-0.51}$ (see Figure 4).
\begin{figure}[t]
 \centering \includegraphics[width=8cm]{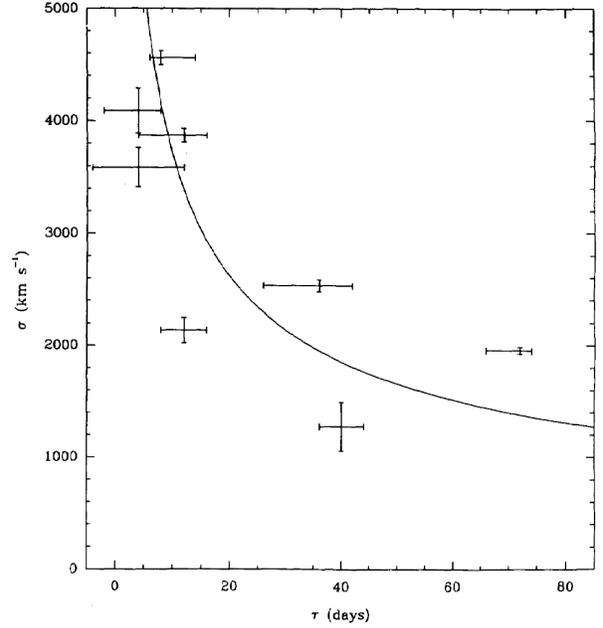}
 \caption{Line width, $\sigma$, versus reverberation mapping time delay, $\tau$, for NGC~5548 from the observations of \citet{Clavel+91}.  The curve is a least-squares fit of $\sigma \propto \tau^{-1/2}$, the expected virial relationship. (Figure adapted from \citealt{Krolik+91}.)}
\end{figure}
The \citet{Krolik+91} result was confirmed for other objects such as NGC~7469 and the radio galaxy 3C 390.3 \citep{Peterson+Wandel00} and the NLS1 Mrk 110 \citep{Kollatschny03}.

As \citet{Krolik+91} pointed out, this correlation between radius and Doppler width is a natural consequence of virialization of all BLR gas motions.  On the other hand, it is {\em very} hard to see how it could arise if the high-ionization lines arise in a wind.  The wind emission would extend through a large range of radii rather than being confined to small radii.  As we will see below when the considering the relationship between the BLR and intrinsic absorption lines, the distances of intrinsic absorption line systems are relatively large (galactic scales).  It should also be noted that the increase in effective radius of the BLR with decreasing ionization does not support a two-component BLR picture (e.g., as proposed by \citealt{Gaskell87} or \citealt{Collin-Souffrin+Lasota88}) since in this picture the high-ionization BLR lines are further out than the low-ionization lines.

\subsection{The line profile}

On the basis of line profiles and object-to-object differences the BLR is sometimes divided into a ``very broad line region'' (VBLR) and a region of lower-velocity gas, sometimes called the ``intermediate-line region´´ (ILR; see \citealt{Baldwin+88}, \citealt{Francis+92}, \citealt{Wills+93} and \citealt{Brotherton+94}). If such a division is real, it is the high-ionization lines of the VBLR which show the blueshifting \citep{Brotherton+94}.  This creates a problem for the outflow model.

Reverberation mapping shows that the high ionization lines come from very close to the centre of the AGN (e.g., \citealt{Krolik+91,Korista+95}). Contrary to what is sometimes depicted in cartoons of AGNs, the accretion disc extends out at least to the radius of the part of the BLR emitting the high-ionization lines.\footnote{The flux at a given frequency comes mostly from the part of a disc at the temperature given by the Stefan-Boltzmann law. The monochromatic luminosity gives the area at around that temperature and hence the size of the disk.  Observationally the size of the disc can be determined from multi-wavelength reverberation mapping \citep{Collier+98,Wanders+97}. It can also be inferred less directly from gravitational microlensing \citep{Mortonson+05}.} The accretion disc is optically thin at optical wavelengths or else it would not produce enough flux.  This means that light cannot pass through it.  In particular, emission-line photons from the far side of the disc cannot pass through it.  Thus in the disc-wind model we should not see any photons coming directly from the redshifted line profiles.  Since we {\em do} see a red side of the VBLR, this is another argument against the high-ionization coming from a wind.

The only way to get {\em any} redshifted emission directly from the disc-wind model is for the outflow to have a large half-opening angle, $\theta$, and to view the system at an inclination, $i$, such that $i > (90^{\circ} - \theta)$.  This is how \citet{Zamanov+02} succeed in approximating an observed blueshifted \ion{C}{4} profile with a disc-wind model.  In the model in their Figure 3 the half-opening angle is $85^{\circ}$ and $i = 15^{\circ}$.  The problem is that the profiles of the Balmer lines constrain $i$ to near face-on values so to satisfy the condition for creating the red side of the line, $\theta$ has to be large.  $\theta = 85^{\circ}$ of the \citet{Zamanov+02} model is problematic because the outflow is skimming the top of the accretion disc by only $5^{\circ}$.  Such an outflow is going to collide at high speed with the dusty torus which is located only just outside the lowest ionization region of the BLR.  It is difficult to see how such a model could work.

We note that a possible solution to the line profile problem is for the redshifted side of a line like \ion{C}{4} to originate from receding gas but seen in reflection from the accretion disc.  This would give us redshifted \ion{C}{4} photons without having to invoke an extreme viewing angle, but the velocity-resolved reverberation mapping problem (see 3.2.1) remains the same.

\subsection{Emissivity}

For any two-body interaction process, such as recombination or collisional excitation, the emissivity per ion is proportional to the density.  For a wind the density falls off rapidly with time and distance.  When the density has fallen by an order of magnitude the emissivity per ion has fallen by an order of magnitude and the contribution to the line emission is negligible.  Gravitationally-bound gas does not have this problem since its density is not falling off with time and distance.

\subsection{Difficulty in explaining the ionization dependence of the blueshift}

As noted above, the blueshifting of high-ionization broad lines increases with ionization.  There is a similar effect for the narrow emission lines.  There is no difficulty explaining this ionization dependence for the {\em narrow} lines because the NLR extends from inside the torus out to many kpc.  It is reasonable to have the far side of the high-ionization and higher velocity inner part of the NLR suffering extinction from the torus while the far side of the more remote, lower velocity/ionization part suffers less extinction.  It is not so easy to apply this explanation to the ionization dependence of the blueshifting of the broad lines because {\em all the BLR comes from within the inner radius of the torus} \citep{Suganuma+06} and so outflowing gas of varying degrees of ionization will {\em all} have the receding far side blocked by the accretion disc and torus.  This is not an insuperable problem, however, because know that the most of the low-ionization BLR emission comes from gravitationally bound gas immediately above the accretion disc.  Thus the increase in blueshifting with ionization could be the result of an increasing wind contribution to higher-ionization lines.

\section{Broad absorption line systems}

A popular early model for the BLR was that it arose in radiation-pressure driven outflow from the AGN (e.g., \citealt{Blumenthal+Mathews75}).  \citet{Collin-Souffrin+88} suggested that the high-ionization BLR clouds were condensations in a wind.  Since blueshifted intrinsic absorption lines are common in AGNs it has been natural to think that the high-ionization BLR, broad absorption lines, and intrinsic narrow absorption lines all originate in a common outflow.  Such a model is discussed, for example, by \citet{Elvis00}.  This model is expanded on by \citet{Risaliti+Elvis10}.  The wind is launched from the surface of the accretion disk and eventually accelerated to the maximum velocities see in BALQSOs ($>> 10^{4}$ km s$^{-1}$).  In this section we point out problems with such models as an explanation of the high-ionization BLR.

\subsection{Observed outflow velocities can greatly exceed BLR velocities}

Velocities in BALQSOs typically range from $0 - 20,000$ km s$^{-1}$, but can be much greater.  For example, SDSS J0230+0059 shows a maximum outflow velocity of $ - 60,000$ km s$^{-1}$ \citep{Rogerson+16}, PG 2302+029 has a maximum velocity of $- 56,000$ km s$^ {-1}$ \citep{Jannuzi+96}, and PG 0935+417 has a maximum velocity of $- 50,000$ km s$^{-1}$ \citep{Rodriguez_Hidalgo+11}.  When we consider the difficulty of detecting broad absorption lines of such extreme velocities it is clear that such high velocities are not rare.\footnote{It should be noted that the SDSS BALQSO catalogue of \cite{Trump+06} is cut off at a maximum outflow velocity of $- 29,000$. This corresponds to the velocity at which \ion{C}{4} absorption can overlap with lower-velocity \ion{Si}{4} absorption.  It is not a real cutoff in the distribution of outflow velocities.}  Such velocities are {\em much} faster than velocities observed in the high-ionization broad lines ($\thickapprox$ 10,000 km s$^{-1}$ -- see, for example, Figure 1 of \citealt{Weymann+81}).  The lack of very high velocity BLR gas suggests that the outflowing gas seen in broad absorption lines does not contribute significantly to the BLR emission.

\subsection{The location and physical conditions of the gas of known outflows are very different the BLR.}

Variability presents additional problems for trying to association high-ionization BLR gas and BAL gas. Variability of BAL lines is common.  They can come and go and change their profiles.  For example, monitoring of SDSS J0230+0059 by \citep{Rogerson+16} shows that the absorption-line profile of the $- 60,000$ km s$^{-1}$ feature is variable on relatively short timescales and a lower velocity, $- 40,000$ km s$^{-1}$, BAL feature appeared during the monitoring.  If such variability is due to changes in ionization, the density is $n_H \sim 10^{5}$ cm$^{-3}$ and the distance of the gas from the centre is in the range of 100 pc to many kpc \citep{Capellup+13,Grier+15,Rogerson+16}.  Such densities and distances are orders of magnitude different from the high densities and small radii of the BLR.

For extensive monitoring of the BALQSO APM 08279+5255, \citet{Saturni+16} have shown that both the BAL absorption and narrow-line absorption are correlated with the continuum variability (thus supporting the hypothesis that EUV variability is a driver of absorption-line variability).  For the narrow-line absorption system they determine a rest-frame time lag of 160 days. This implies a density of $2 \times 10^{4}$ cm$^{-3}$ and a distance of the order of 10 kpc.  For the BAL system the most likely time lag is 20 times smaller implying a density 20 times higher, but the current uncertainty in the lag is too large to draw conclusions.

Another variability model considered by (see discussion in \citealt{Rogerson+16}) is the ``flow-tube'' model.  In this the BAL gas flows through ``tubes'' which in addition to having outflowing gas also have a transverse component of velocity.  It is this transverse velocity of the tube which is responsible for the observed variability.  Clearly, if this model is correct, the motions of the gas are very different from what is needed for the BLR.

\subsection{Broad absorption frequently occurs at zero or even positive relative velocities}

It must be remember that, by definition, BALs are {\em broad}, i.e., absorption is seen over a wide {\em range} of velocities. A BAL outflow is almost certainly accelerating, meaning that the {\em lowest} outflow velocities are closest to the centre.  Reverberation mapping shows that for the BLR it is {\em high} velocity gas that is close to the centre.  This situation gets worse when we consider what the minimum BAL velocities are.  While the maximum velocity of outflows can be very high ($\sim 0.2c$, see above), the most common {\em minimum} outflow velocity of BAL troughs is {\em zero} (see, for example, Figure 1 of \citealt{Lee+Turnshek95}).  Furthermore, there are also clear cases of {\em redshifted} broad absorption lines, i.e., {\em inflowing} BAL gas \citep{Hall+13}.  These cases are not simply due to underestimating the redshift of the AGN because of blueshifting of high-ionization lines \citep{Gaskell83}, but are true broad absorption systems (trough widths $\Delta v > 3000$ km s$^{-1}$) with maximum inflow velocities of up to $v \thickapprox +12,000$ km s$^{-1}$.  These AGNs mostly also show blueshifted absorption. The commonness of absorption at $v \sim 0$ and the existence of redshifted BAL systems argues against the na{\"i}ve model of intrinsic absorption line originating in an outflow with the high-ionization BLR.

\subsection{The covering factor of high-ionization broad emission lines is high, but not every AGN has broad absorption lines.}

From energetics considerations the covering factor of \ion{C}{4} emission has to be very high \citep{Gaskell+07}.  This is also supported by reverberation mapping \citep{Gaskell+07, Kollatschny+Zetzl13}.  If the high-ionization BLR were outflowing we would therefore expect essentially {\em every} AGN to show blueshifted \ion{C}{4} absorption.  Since this is not the case, particularly for low-luminosity AGNs, this is another argument against all the high-ionization BLR emission arising from an outflow.

\subsection{BALs show strong redshift evolution, but the blueshifting of high-ionization emission lines does not.}

The incidence of BAL systems increases strongly with redshift.  \citet{Allen+11} report a strong factor of $3.5 \pm 0.4$ decrease in the incidence of BAL systems from redshifts $z \thickapprox 4.0$ down to $z \thickapprox 2.0$.  This suggests that the presence of BALs is associated with the nuclear environment rather than being associated with the BLR.  It also suggests that the BAL phenomenon is unrelated to the blueshifting of high ionization lines because, as noted above, the blueshifting does not depend on luminosity or redshift.

\section{The scattering + inflow model}

We have argued \citep{Gaskell+Goosmann13} that rather than being due to occultation of the receding side of an outflow, the blueshifting is due to {\em scattering} off an inflowing medium. Electron scattering as a possible factor causing line broadening in AGNs is an old idea \citep{Kaneko+Ohtani68,Weymann70,Mathis70}. If there is a component of net radial motion this will produce a shift of the line profile (e.g.,
\citealt{Auer+Vanblerkom72}).  The reason is explained in Figure 5.
%
\begin{figure}[t]
 \vspace{5pt}
 \centering \includegraphics[width=8cm]{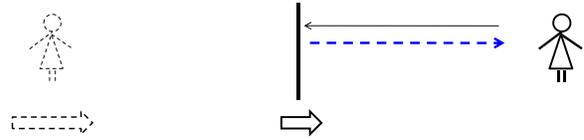}
 \caption{Illustration of why scattering off a medium with a net inflow produces a blueshift.  If a mirror is moving towards someone then the image in the mirror is approaching at twice the velocity of the mirror and hence the light is blueshifted. (Figure from \citealt{Gaskell09}.)}
\end{figure}
The simulations of \citet{Kallman+Krolik86} and \citet{Ferrara+Pietrini93} show such shifts and it
has been suggested that this could be a cause of the blueshifting of high-ionization lines \citep{Corbin90,Mathews93}.  As \citet{Korista+Ferland98} point out, the optical and UV albedo of BLR clouds can be significant because of Rayleigh scattering in addition to Thomson scattering.

\subsection{Line profiles modified by scattering in the presence of a net inflow}

We have used the {\it STOKES} Monte Carlo radiative transfer code \citep{Goosmann+Gaskell07,Marin+12,Marin+15} to investigate line profiles produced when there is a net inflow of the BLR.  We only describe some of the results here.  More details can be found in \citet{Gaskell+Goosmann13}.  Our main result is that we can readily reproduce the observed range of profiles and this result is insensitive to the precise geometry.  The insensitivity to geometry is because the observed  profile is the sum of the unscattered line profile and the line profile for the scattered light. For a high covering factor the ease with which the original photons can escape depends on the optical depth.  For low covering factor it depends more on the covering factor. The blueshifting of the scattered light depends on the inflow velocity of scattering material immediately outside the emitting region.  The blueshift of the resulting observed profile thus depends on both the inflow velocity and the ease of escape of photons.

In Figure 6 we show how the predicted line profiles vary with the optical depth of the scattering medium for a high covering factor.  We have assumed the limiting case of a 100\% covering factor.  It can be seen that there is good agreement with the observed composite profiles shown in Figure 1.

\begin{figure}[t]
\vspace{1pt}
\centering \includegraphics[width=8cm]{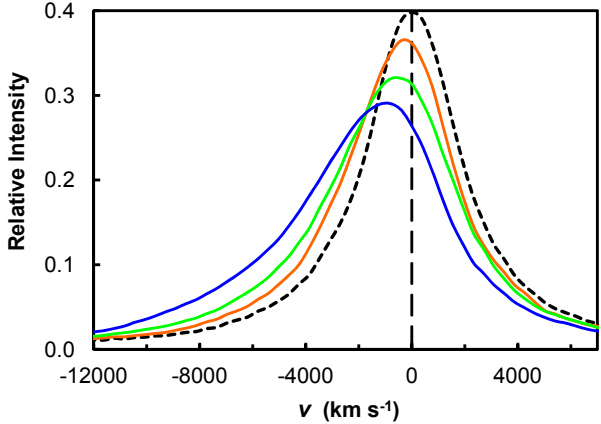}
\vspace{10pt}
 \caption{Calculated shifts of the \ion{C}{4} line by electron or Rayleigh
scattering. The dashed black curve shows a Lorentzian line
profile before scattering. The other solid curves show the
blueshifting caused by a spherical distribution of scatterers
with an inflow velocity of 1000 km s$^{-1}$. The curves are normalized to the same total flux (i.e., the same area).  In order of increasing blueshifting and
decreasing peak flux, the curves show the resulting profiles for optical depths of $\tau$ = 0.5
(red), 1.0 (green), and 2.0 (blue). The predicted curves can be compared with observed ones
in Figure 1.}
\end{figure}
%
\begin{figure}[h]
\vspace{1pt}
\centering \includegraphics[width=8cm]{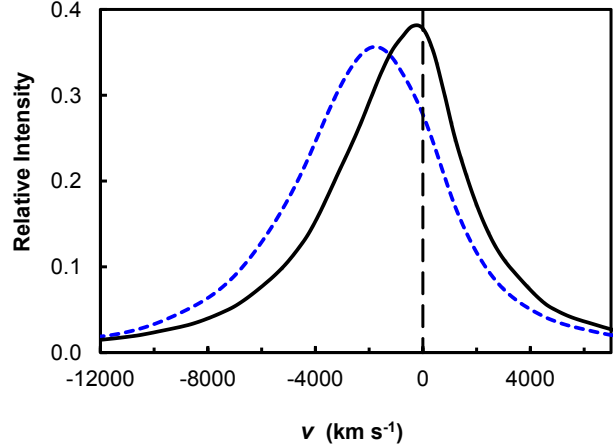}
\vspace{10pt}
 \caption{Predicted profiles for a flared equatorial disc of inflowing scatterers with half-opening angle of $60^{\circ}$ and a radial optical depth of $\tau_{rad} = 5$. The solid black curve shows the profile
predicted when the system is viewed near face on. The
dashed blue line shows the profile predicted when the system is viewed from
just inside the disk surface as could be the case in a BALQSO.  The unscattered line profile is as in Figure 6. (Figure from \citealt{Gaskell+Goosmann13}.)}
\end{figure}

In Figure 7 we show the predicted profiles for a more likely equatorial distribution of inflowing scatterers.  We have taken the half-opening angle of the flared disc of scatters to be $60^{\circ}$.  The resulting profiles are not very different from Figure 6.  In this case the viewing angle also matters. When the system is viewed from just inside the scattering disc the profile shift is larger and the line somewhat broader.  This could be the case for a BALQSO.  As noted above, BALQSOs show larger blueshifts of \ion{C}{4}.  We have also noted \citep{Gaskell+Goosmann13} that more blueshifted \ion{C}{4} lines are broader.  An important thing to note from comparing Figures 6 and 7 is that the profiles arising from distributions of scatterers with quite different geometries are similar.

In Figure 8 we show comparisons of our theoretical profiles to the observed profile of a single object.  Note that comparable fits can be obtained for different geometries because the profiles are insensitive to the geometry.  There is a degeneracy between the the effect of the ease of escape of the photons and a different inflow velocity.
\begin{figure}[t]
\vspace*{0.3cm}
\centering \includegraphics[width=84mm]{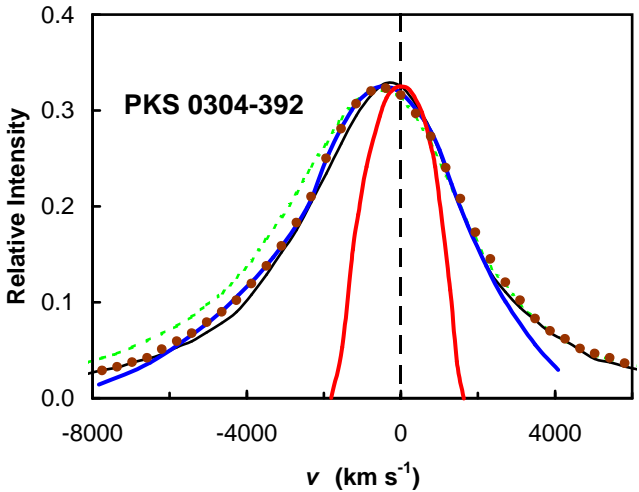}
\caption{The profiles of O I $\lambda$1305 (narrow symmetric profile
shown in red) and \ion{C}{4} $\lambda$1549 (thick blue line) for the
quasar PKS~0304-392.  The thin black line is the blueshifted profile
produced by a spherical distribution of scatters with $\tau$ = 0.5,
and the dashed green line is the profile produced by the same
distribution with $\tau$ = 1.  The brown dots are the profile
produced by a $\tau$ = 20 inflowing cylindrical distribution. In both cases the inflow velocity is 1000 km s$^{-1}$.
(Figure from \citealt{Gaskell+Goosmann13}; PKS~0304-392 observations taken from \citet{Wilkes84}.)} \label{PKS_fit}
\end{figure}

\subsection{Velocity-dependent time delays produced by scattering}

The scattered line photons have a slightly greater time delay than the line photons seen directly.  This is because the photons have a longer path to travel.  The extra path is comparable to the size of the region emitting the line.  Because light of all velocities is scattered the inflow+scattering model will produce a fairly symmetric velocity-dependent time delay with slightly greater delays for the blue wing because of the net blueshifting of the scattered light.  This sort of velocity dependence of the lags is commonly seen. (see, for example, Figure 5c of \citealt{DeRosa+14})

\subsection{Dependence of blueshifting on radius of emission}

It is well established, via size estimates from reverberation mapping and because of the dependence of line width on ionization, that lines of different ionizations are produced at different radii, $R$ (see section 5.2 above).  Because inflow is driven by gravity, the velocity of inflow will increase with decreasing radius as $v_{inflow} \propto R^{-1/2}$.  In the infall+scattering model the blueshifting of a given emission line depends on the velocity of inflow just outside the line-emitting region.  The model therefore predicts that the blueshifting of a line should go as $R^{-1/2}$.  As can be seen in Figure 9, this indeed seems to be the case.
\begin{figure}[t]
\vspace*{0.3cm}
\centering \includegraphics[width=84mm]{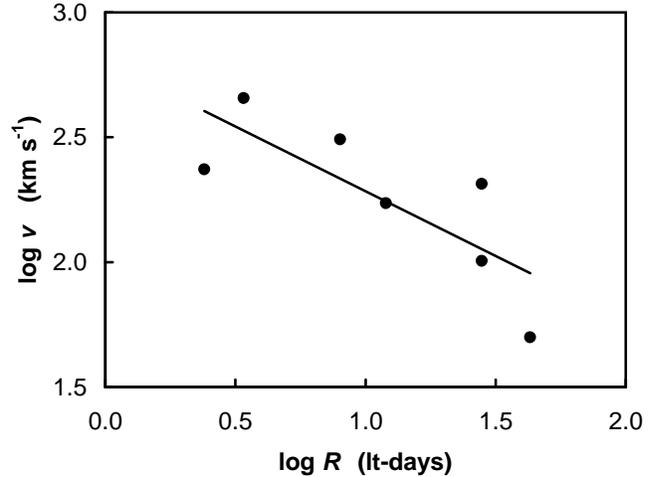}
\caption{Mean blueshifts of emission lines (see \citealt{Gaskell+Goosmann13} for details)
versus the radii, $R$, of maximum emission. The line is a least squares fit showing $\log v$ as a
function of $\log R$. (Figure from \citet{Gaskell+Goosmann13}.} \label{relative_shifts}
\end{figure}

\section{Additional reverberation mapping support for inflow of the BLR}

As discussed above (see 5.1), velocity-resolved reverberation mapping has long excluded outflow of high-ionization lines and favoured a modest inflow \citep{Gaskell88,Koratkar+Gaskell89}. From available UV reverberation mapping \citet{Gaskell+Snedden97} concluded that for essentially all AGNs with suitable data, profile variations of the C IV line were consistent with a slight, but significant net inflow in all cases.  These early studies were mostly simply looking at the delays of the red side of the line compared with the blue side of the line.  With more intensive monitoring it is possible to get better velocity resolution and eventually to recover the transfer function (the response of a system to a $\delta$-function in the driving time series) as a function of velocity and hence to produce three-dimensional ``velocity-delay maps''.

The first of 3-D velocity-delay map to be produced \citep{Ulrich+Horne96} confirmed that the variability of \ion{C}{4} in NGC~4151 was {\em not} consistent with \ion{C}{4} being produced in a wind, but instead favoured virialization.  \citet{Korista+95} carried out {\it HST} monitoring of NGC~5548.  The \ion{C}{4} profile of NGC~5548 shows a clear blueshift.  The \ion{C}{4} profile for the low state in the 1989 campaign of \citet{Clavel+91} is shown in Figure 10. A similar blueshift was also present during the {\it HST} campaigns of 1993 and 2013 as can be seen in Figure 7 of \citet{Korista+95} and Figure 5 of \citet{DeRosa+14}.
%
\begin{figure}[t]
\vspace*{0.3cm}
\centering \includegraphics[width=80mm]{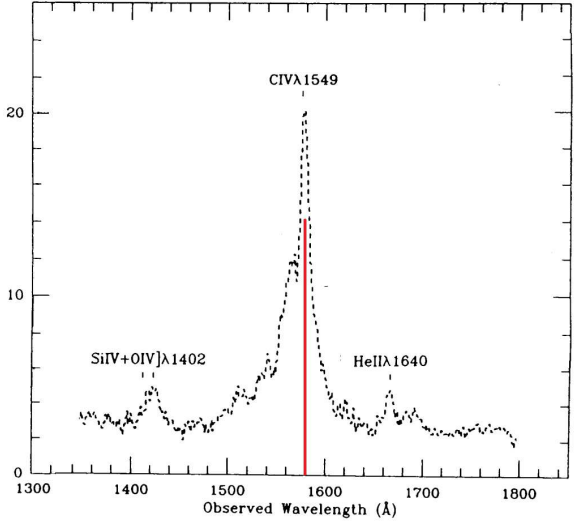}
\caption{The profile of \ion{C}{4} during the low state of the 1989 campaign.  The red line indicates the peak of the narrow component. (Figure adapted from \citealt{Clavel+91}.)}
\end{figure}
If this blueshifting is due to a wind, NGC~5548 would therefore be a good candidate for showing the velocity-resolved variability signature of a wind (i.e., the blue wing leading the red wing).  However, the analyses of \citet{Kollatschny+Dietrich96} and \citet{Done+Krolik96} for the 1989 and 1993 campaigns respectively show the {\em red} wing leading.  The quantitative modelling of \citet{Done+Krolik96} gives an inflow velocity of $\thickapprox - 1400$ to $- 2000$ km s$^{-1}$.  This is supported by the 3-D velocity-delay map from the 1993 {\it Hubble Space Telescope} ({\it HST}) observations (see Figure 37 of \citealt{Peterson01}).  {\it HST} monitoring of NGC~5548 has been repeated two decades later by \citet{DeRosa+14} with an essentially identical result to the 1989 and 1993 campaigns: the lags of the red side of \ion{C}{4} lead the lags of the blue side (see their Figure 5c).

Over the past decade large amounts of ground-based telescope time have been invested in obtaining vastly better data sets and more sophisticated analyzes have been carried out. The best studied line is H$\beta$.  From their velocity-resolved reverberation mapping \citet{Grier+13} find H$\beta$ to be inflowing in each of the five diverse AGNs they studied (Mrk~6, Mrk~335, Mrk~1501, PG 2130+099, and 3C 120).  From a study of another five AGNs (Mrk 40 = Arp 151, Mrk 1310, NGC 5548, NGC 6814, and SBS 1116+583A), \citet{Pancoast+14} conclude that four out of the five have a net inflow and one case (the NLS1 Mrk 1310) is ambiguous.

NGC~5548 is unfortunately the only AGN for which {\it HST} monitoring of \ion{C}{4} has successfully been carried out.  However, several ground-based monitoring campaigns have produced 3-D velocity-delay maps of \ion{He}{2} $\lambda$4686, a line coming from even higher ionization gas than \ion{C}{4}.  The first was the {\it Hobby-Ebberly Telescope} monitoring of the NLS1 Mrk~110 by \citet{Kollatschny+01}.  This is shown in Figure 11.  It can be seen that this also rules out an outflowing wind contribution.  Instead, the red side of \ion{He}{2} slightly leads the blue side (i.e., the lag at positive velocities is slightly less).  This favours bound gas with a slight net inflow.
\begin{figure}[t]
\vspace*{0.3cm}
\centering \includegraphics[width=84mm]{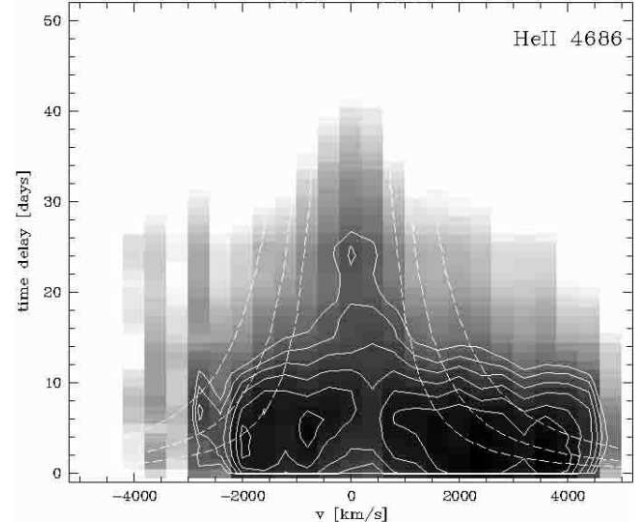}
\caption{\ion{He}{2} velocity-delay maps for the NLS1 Mrk~110.  The dotted lines show the $\tau \propto v^{-2}$ outer envelopes expected from an infinite, virialized, distribution of emitting gas for different black hole masses (see \citealt{Kollatschny03}.)   (Figure reproduced from \citealt{Kollatschny03}.)}
\end{figure}

\citet{Grier+13} present 3-D velocity-delay maps for \ion{He}{2} $\lambda$4686 lines of an additional three AGNs. These also fail to show an outflow signature.  The case of Mrk~335 is particularly interesting.  Mrk~335 is a NLS1 and, as noted earlier, NLS1s show the largest blueshifts of high-ionization lines.  Both the \ion{C}{4} and Lyman $\alpha$ line profiles show a blueshifting (see Figure 12).  Although this is not as extreme as for some NLS1s, this places Mrk 335 in the second highest group of blueshifted profiles in Figure 1.
\begin{figure}[t]
\vspace*{0.3cm}
\centering \includegraphics[width=84mm]{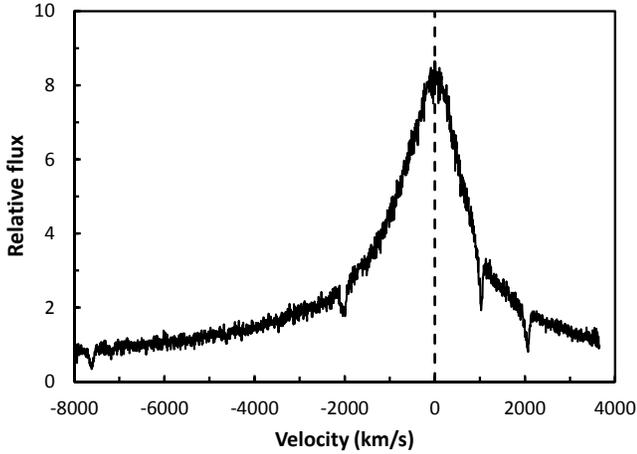}
\caption{High-resolution Ly $\alpha$ profile of Mrk~335.}
\end{figure}
Despite this, the 3-D delay map of the \ion{He}{2} $\lambda$4686 line (see Figure 13) shows no evidence for a wind.  \citet{Grier+13} conclude instead that the \ion{He}{2} emission is consistent with an inclined disc.
\begin{figure}[t]
\vspace*{0.3cm}
\centering \includegraphics[width=65mm]{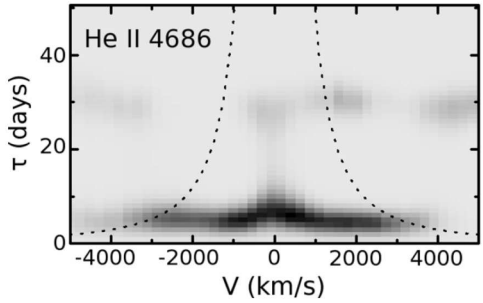}
\caption{\ion{He}{2} velocity-delay map for the NLS1 Mrk~335.  The spurious horizontal features at a lag of $\sim 30$ days are the result of aliasing. The dashed curve gives the predicted envelope for the delays for an infinite distribution of virialized gas around a $4.6 \times 10^{6}$ solar mass black hole. (Figure from \citealt{Grier+13}.)}
\end{figure}

\section{Discussion}

\subsection{Cases of apparent outflow}

Although the vast majority of reverberation mapping studies point to a net inflow of the BLR (see above), there have been a couple of observations indicating outflow.  The first of these was found in the \citet{Kollatschny+Dietrich96} analysis of the \ion{C}{4} profiles during the 1989 NGC~5548 campaign \citep{Clavel+91}.  As noted above (see section 5.1 and 8), for the campaign as a whole the red side of \ion{C}{4} leads the blue side.  However, if one looks at the lags for the first and second events of the 1989 campaign separately (see Figure 14) one sees that during the first event {\em the blue side of the profile varied first}.  Taken at face value these velocity-dependent lags would imply an outflowing BLR during the first event and an inflowing BLR during the second event.  Obviously this cannot be right because a BLR cannot suddenly change its direction in only a couple of months!  \citet{Gaskell10,Gaskell11} proposed instead that the change in apparent kinematic signature on such a short timescale was due to changes in the illumination of the BLR by off-axis flaring regions \citep{Gaskell08}.  The first event would be caused by a flaring region on the approaching side of the disc and the second event by one on the receding side of the disc.  \citet{Gaskell10} shows that the apparent outflow inferred by \citet{Denney+09} from their intensive 2007 campaign of monitoring H$\beta$ variability of NGC~3227 can readily be explained in this manner.

\begin{figure}[t]
 \vspace{5pt}
 \centering \includegraphics[width=6.5cm]{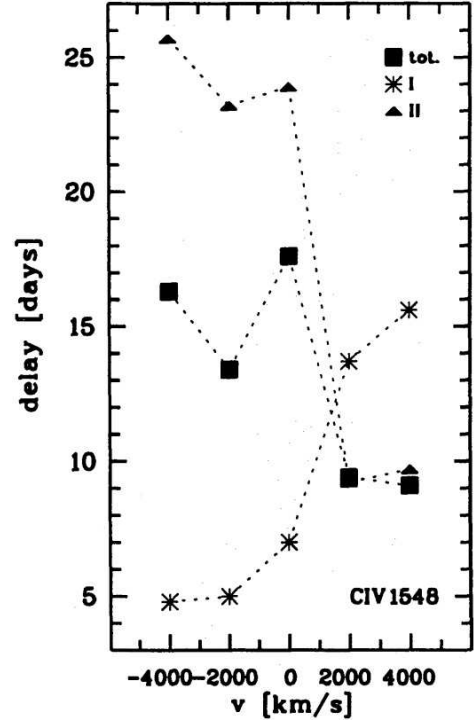}
 \caption{Velocity-dependent lags for \ion{C}{4} for NGC~5548 during the 1989 observing campaign. The squares are the lags for the entire campaign, the asterisks for the first event, and the triangles for the second event.  (Figure adapted from \citealt{Kollatschny+Dietrich96}.)}
\end{figure}

There have been some cases of apparent strong inflows of the H$\beta$-emitting regions such of Mrk~6 \citep{Sergeev+99} and Mrk~40 \citep{Bentz+08}.  These strong inflows can be explained in the same way as the apparent outflow of the BLR of NGC~3227 in 2007.  If the central flare is displaced toward the receding side of the disc the red profile of the line will respond first.  As shown in \citet{Gaskell10}, the velocity dependent lags of these cases of apparently strong inflow look the same as the lags for the apparent outflow of Mrk~40 if one merely reverses the sign of the velocities.

\subsection{Does optical \ion{Fe}{2} emission show an inflow?}

\citet{Hu+08} have claimed that there is a redshifting of the optical \ion{Fe}{2} $\lambda$4570 emission blend relative to the NLR and H$\beta$. They adopt a similar explanation to that proposed by \citet{Gaskell82} for the blueshifting of high ionization lines -- i.e., that the far side of the emission is blocked by the accretion disc and torus.  To explain a redshifting, however, \citet{Hu+08} propose that the region producing the optical \ion{Fe}{2} is {\em inflowing}. They claim inflow velocities of  $\thickapprox 400$ km s$^{-1}$ and going up to 2000 km s$^{-1}$.  Such redshifts are unlikely to be real since the study of \citet{Phillips78a,Phillips78b}, which used much higher quality data than the SDSS spectra analyzed by \cite{Hu+08} and which did a careful de-blending of individual \ion{Fe}{2} lines with multiplets, did not find any such shift of the \ion{Fe}{2} lines.  Phillips discussed the already then well known blueshifts of the high-ionization narrow lines, but says nothing about any redshifting of the \ion{Fe}{2} lines.   \citet{Sulentic+12} created high signal-to-noise composites of the same SDSS spectra used by \citet{Hu+08} and additional AGNs and found no evidence for a systematic redshift.  As \citet{Sulentic+12} note, \citet{Hu+08} did not subtract out the \ion{He}{2} $\lambda$4686 emission and this is the probable cause of the spurious \ion{Fe}{2} redshifts claimed by the latter.

\subsection{The correlation of broad-line and narrow-line blueshiftings}

The correlation of broad-line and narrow-line blueshiftings is certainly a good indirect argument for the blueshifting of the high-ionization broad lines being due to a wind.  If the inflow+scattering model is correct instead then the correlation between the NLR and BLR blueshifting must be due to a common factor.  This factor is most likely to be the Eddington ratio.  A high Eddington ratio means a high relative accretion rate and, as noted, this will cause an increased blueshift in the inflow+scattering model because of the higher inflow velocity and higher column density.  A high Eddington ratio can also drive stronger NLR outflow.  This can happen in a couple of ways.  A higher inflow rate requires an increased rate of angular momentum removal so there has to be a stronger wind.  The increased radiation pressure compared with gravity for a higher Eddington ratio means that the NLR gas can be driven out more efficiently.

\subsection{Is there a redshifting of H$\beta$?}

In their principal component analysis \citet{Boroson+Green92} found the apparent asymmetry of H$\beta$ to be a strong component of their Eigenvector 1 (EV1).  The correlation with the strength of [\ion{O}{3}]~$\lambda$5007 is particularly striking.  This is shown in Figure 15.  It can readily be seen that weak [\ion{O}{3}] objects (e.g., NLS1s) show a blueward asymmetry of H$\beta$, whereas strong [\ion{O}{3}] objects seem to have a redward asymmetry.  From a comparison of \ion{C}{4} and H$\beta$ \citet{Sulentic+95} found that the objects with a strong apparent redshift of H$\beta$ did not show a blueshifting of \ion{C}{4} and vice versa (see their Figure 1).
\begin{figure}[h]
 \vspace{5pt}
 \centering \includegraphics[width=8cm]{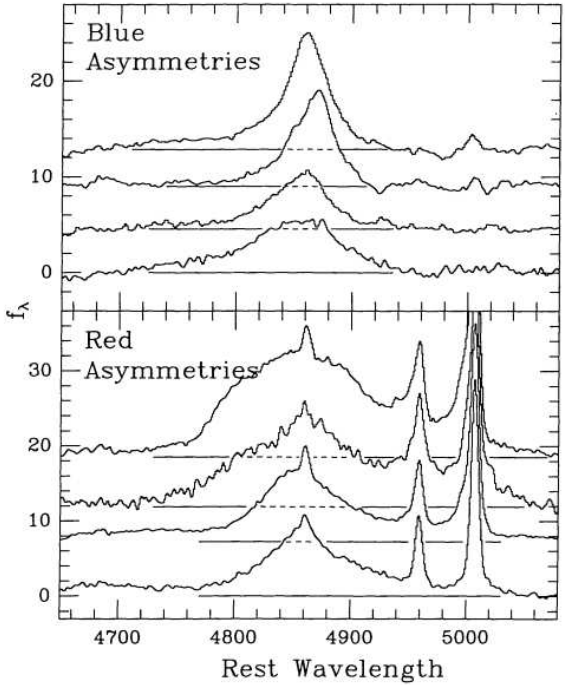}
 \caption{H$\beta$ profiles for AGNs with weak [\ion{O}{3}] (top) and strong [\ion{O}{3}] (bottom). The extent of the wings of H$\beta$ is indicated by the extent of the horizontal zero line for each object.  \ion{Fe}{2} emission has been subtracted out.  (Figure from \citealt{Boroson+Green92}.)}
\end{figure}

A redshifting of H$\beta$ that is anti-correlated with the blueshifting of \ion{C}{4} is hard to understand in both the disc-wind model and the inflow+scattering model.  If the redshifting of H$\beta$ is due to the accretion disc blocking the far side of the BLR then the gas producing H$\beta$ must be inflowing.  Inflow would be expected to be greatest in objects with high Eddington ratios, but these are the very objects that do {\em not} show the apparent H$\beta$ redshifting.  If the redshifting of H$\beta$ if due to scattering then the gas producing H$\beta$ must be outflowing.  Then one wonders why this gas, which is associated with lower-ionization gas such as that producing \ion{Mg}{2} and \ion{Fe}{2}, is outflowing more than the high-ionization gas.

We propose that the solution to this problem is that the apparent redshifting of H$\beta$ is not real but is due to is due to extended wings of [\ion{O}{3}].  This is not a new suggestion.  It was recognized in the 1980s that the apparent redward asymmetry of H$\beta$ (commonly referred to in the literature of the time as ``the red shelf'') is not a real H$\beta$ asymmetry because H$\alpha$ and H$\gamma$ do not show the same asymmetry (see \citealt{Osterbrock+Shuder82}).  The asymmetry must therefore be due to blending with other lines.  There was some debate in the literature over the relative contributions of different ions to the red shelf.  \citet{Osterbrock+Shuder82} attributed the red shelf to \ion{Fe}{2}.  This is strongly supported by the good correlation between the strength of the red shelf with other optical \ion{Fe}{2} emission (see Figure 2 of \citealt{Joly88}) and by correlated line variability \citep{Doroshenko+99}.  \citet{Jackson+Browne89} found red shelves to be common among radio-loud AGNs which often do not show strong optical \ion{Fe}{2}.  They suggested that weak \ion{He}{1} lines could be a contributor to the shelf in objects where \ion{Fe}{2} was weak. From the shape of the red shelf in Ark~120 \citet{Foltz+83} argued that, in at least that AGN, broad [\ion{O}{3}] was a better fit to the shelf the \ion{Fe}{2}.  \citet{Meyers+Peterson85} suggested contributions from both \ion{Fe}{2} and broad [\ion{O}{3}].

For objects with significant \ion{Fe}{2} emission most of the red shelf can be successfully be removed using a template of \ion{Fe}{2} emission (see, for example, Figure 15).  What does not seem to have  been recognized is that when the NLR is strong, a residual shelf remains after \ion{Fe}{2} emission has been subtracted out.  This can be seen in Figure 15.   Variability of Fairall 9 provided strong support for the residual red shelf being due to the NLR because there was a residual constant component when the broad lines varied \citep{Stirpe+89}.  This is mostly due to broad wings of [\ion{O}{3}] but \citet{Veron+02} found that various other weaker narrow lines make a non-negligible contribution to the red shelf.

We believe that an [\ion{O}{3}] origin of the residual red shelf of H$\beta$ explains the correlation of the apparent asymmetry of H$\beta$ with other properties.  The measured asymmetry is part of EV1 simply because EV1 is dominated by the strength of [\ion{O}{3}] and the apparent redward shift of H$\beta$ in objects is caused by H$\beta$.  The relationship between the apparent velocity shift of H$\beta$ and the blueshifting of the high-ionization lines is a consequence of the correlation of the latter with EV1.

\section{Conclusions}

Although theoretically attractive, the disc-wind explanation of the blueshifting of high-ionization broad lines has a number of problems.  Foremost among these is the failure of velocity-resolved reverberation mapping to show the expected kinematic signature of an outflowing wind.  Instead, line profile variability indicates a net inflow of the BLR.  This already strong conclusion has been further reinforced by the new generation of reverberation-mapping studies.  The scattering+inflow model avoids the problems of the disc-wind model.  The scattering+inflow model is insensitive to the geometry and does not require any fine tuning.  Covering factors in AGNs are large and albedos are not zero.  Therefore scattering must occur.  Thus, if there is an inflowing component of the BLR velocity field, scattering will cause a blueshifting to occur.  There certainly must be winds in AGNs, but the idea that they contribute significantly to the emission of high-ionization lines runs into major difficulties.

\acknowledgments

We are grateful to Jack Sulentic, Paola Marziani, and Ski Antonucci for useful discussion and to Todd Boroson, Richard Green, Kate Grier, Wolfram Kollatschny, and Julian Krolik for permission to use figures from their papers. We would also like to express our appreciation to the anonymous referee for a detailed and helpful report.  RWG acknowledges support from French grant ANR-11-JS56-013-01.


%

\end{document}